\documentclass[aps,prl,twocolumn,superscriptaddress]{revtex4-1}
\usepackage{graphicx}

\begin{document}

\title{Length Scale of Correlated Dynamics in Ultra-thin Molecular Glasses}
\author{Yue Zhang}
\author{Ethan Glor}
\affiliation{Department of Chemistry, University of Pennsylvania.}
\author{Mu Li}
\affiliation{Department of Chemistry, University of Pennsylvania.}
\affiliation{Department of Materials Science and Engineering, University of Pennsylvania.}
\author{Tianyi Liu}
\author{Kareem Wahid}
\author{William Zhang}
\affiliation{Department of Chemistry, University of Pennsylvania.}
\author{Robert Riggleman}
\affiliation{Department of Chemical and Bimolecular Engineering, University of Pennsylvania}
\author{Zahra Fakhraai}
\affiliation{Department of Chemistry, University of Pennsylvania.}
\date{\today}

\begin{abstract}
Physical vapor deposition (PVD) is widely used in manufacturing ultra-thin layers of amorphous organic solids. Here, we demonstrate that these films exhibit a sharp transition from glassy solid to liquid-like behavior with thickness below 30~nm. This liquid-like behavior persists even at temperatures well below the glass transition temperature, T$_{\mathrm{g}}$. The enhanced dynamics in these films can produce large scale morphological features during PVD and lead to a dewetting instability in films held at temperatures as low as T$_{\mathrm{g}}$-35~K. We measure the effective viscosity of organic glass films by monitoring the dewetting kinetics. These measurements combined with cooling rate-dependent T$_{\mathrm{g}}$ measurements show that the apparent activation barrier for rearrangement decreases sharply in films thinner than 30~nm. These observations suggest long-range facilitation of dynamics induced by the free surface, with dramatic effects on the properties of nano-scale amorphous materials.
\end{abstract}

\maketitle

Nanometer-sized thin films of small organic molecules are widely used in applications ranging from organic photovoltaics\cite{Shirota2007} and light emitting diodes\cite{Veinot2005,Tang1987}, to protective coatings\cite{Mahltig2005} and high resolution nano-imprint lithography\cite{Pires2010}. 
It is advantageous to use amorphous films because, compared to crystals, they do not have grain boundaries to hinder charge transport, generate cracks and defects, or disrupt the writing processes. 
Physical vapor deposition (PVD), the common method used to manufacture these films, is usually performed at substrate temperatures below T$_{\mathrm{g}}$ to produce films in the glassy state. However, if the properties at nanoscale deviate significantly from bulk properties, the resulting films can have reduced kinetic and thermal stability. Recent experiments suggest that diffusion at the free surface of organic glasses can be several orders of magnitude faster\cite{Zhu2011,Daley2012}, with weaker temperature dependence compared to bulk diffusion. Enhanced, weakly temperature-dependent dynamics on the surface of polymeric glasses\cite{Fakhraai2008,Paeng2011} have been shown to significantly affect the properties of ultra-thin polymer films\cite{Keddie1994,Ediger2014,Ellison2003,Shavit2013,Glor2014,Yang2010,Paeng2011,Pye2010,Priestley2005}. 
In polymeric systems, the molecular weight of the polymer\cite{Glor2014}, and the temperature range of the measurement\cite{Fakhraai2008,Paeng2011,Glor2014} seem to also affect the observed properties, resulting in ambiguity in the relationship between enhanced dynamics at the free surface and properties of ultra-thin glass films. As such, these results can not be extrapolated to molecular and atomic glass systems. 

To our knowledge, there are no systematic studies that measure the dynamics of ultra-thin films of organic glasses with thicknesses less than 100 nm. However, indirect evidence suggests that the properties of these films may be strongly thickness dependent\cite{Sepulveda2011}. Thick PVD films have been shown to form exceptionally stable glasses upon deposition at temperatures just below T$_{\mathrm{g}}$\cite{Swallen2007,Leon-Gutierrez2010,Fakhraai2011,Liu2015a}. While the detailed mechanisms of the formation of stable PVD glasses are still under investigation, most studies \cite{Swallen2007,Leon-Gutierrez2010,Chua2015,Singh2011a} indicate that surface-mediated equilibration (SME) is critical to their production. As such, it is imperative to study the extent and the length scales of the effect of enhanced surface mobility on the dynamics of ultra-thin films. Furthermore, these length scales may also be directly compared with fundamental length scales of glass transitions as proposed by various theories\cite{PazminoBetancourt2015,Mirigian2014,Hocky2012,Butler1991}. 

In this letter, we use dewetting kinetics to measure the dynamics of ultra-thin films of the molecular organic glass, N,N$^{'}$-Bis(3-methylphenyl)-N,N$^{'}$-diphenylbenzidine (TPD). While a direct measure of absolute viscosity of thin films can not be obtained due to potential gradient in the dynamics induced by free interface, by relating the dewetting times with cooling rate-dependent T$_{\mathrm{g}}$ (CR-T$_{\mathrm{g}}$) measurements\cite{Glor2014,Glor2015}, we are able to measure the \lq\lq{effective viscosity\rq\rq} of ultra-thin films as function of film thickness and temperature. In the absence of gradients in the dynamics, the effective viscosity equals the film viscosity. We show that ultra-thin films remain mobile far below bulk T$_{\mathrm{g}}$, and the apparent activation energy for dewetting decreasing sharply for film thicknesses below 30 nm. 

\begin{figure}[b]
	\includegraphics[width=.4\textwidth]{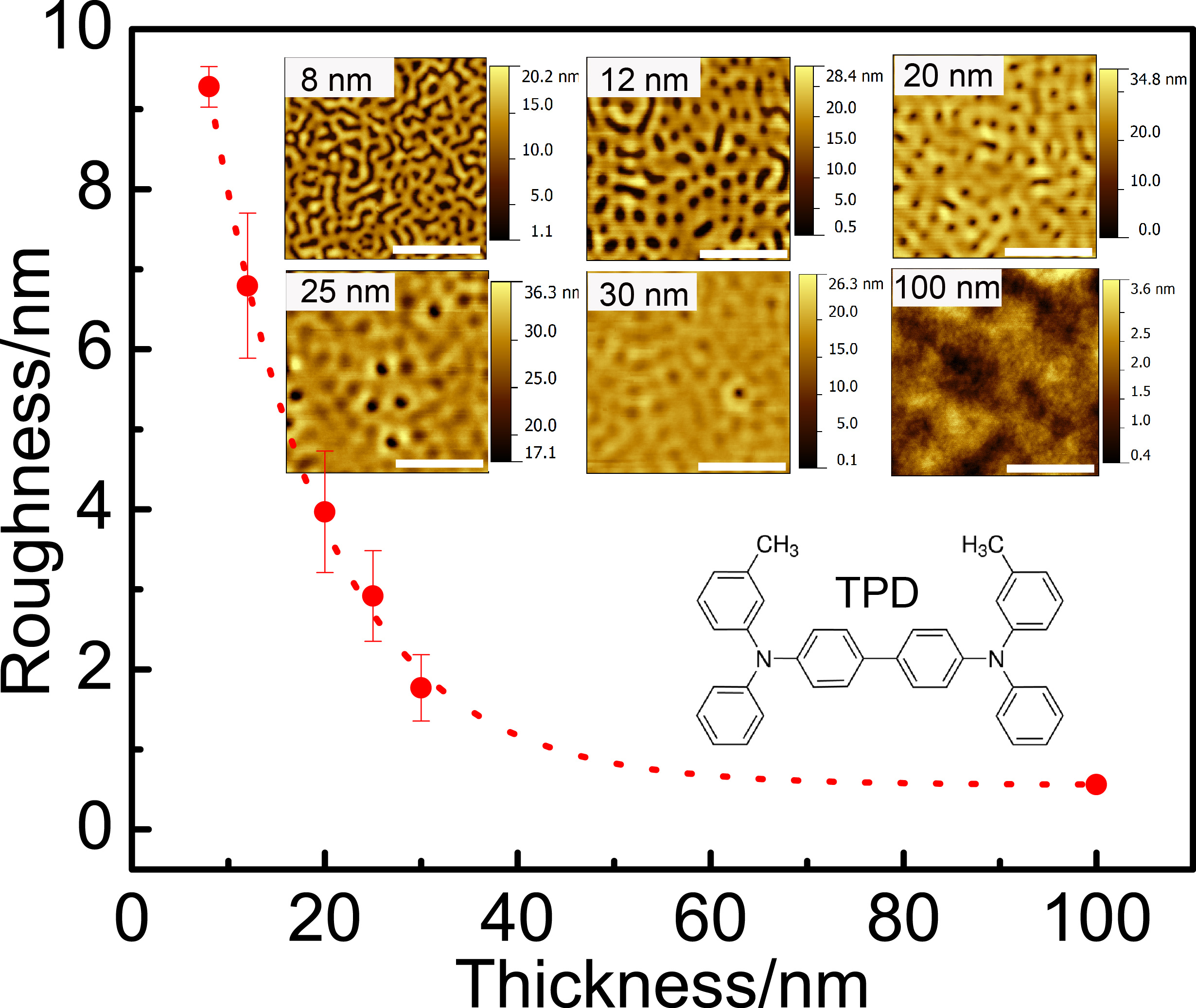}
	\caption{Root-mean-square (RMS) roughness of as-deposited films as a function of film thickness. All films were deposited at T$_{\mathrm{g}}$ with a deposition rate of 0.02~nm/s. Insets show representative AFM images of the as-deposited films from which roughness were calculated. Scale bar in each image is 2~$\mu$m. Molecular structure of TPD is shown as inset.}
\end{figure}

Thin TPD films were prepared by PVD under ultra-high vacuum conditions to ensure uniform substrate properties (Details in SI and Fig. S3). Fig. 1 shows the root mean square (RMS) roughness of as-deposited films produced at a deposition rate of 0.02~nm/s, and a substrate temperature of bulk T$_{\mathrm{g}}$ (328~K) (Material characterization in Fig. S1). The insets show representative atomic force microscopy (AFM) images of as-deposited morphologies, typically imaged within 15 minutes of deposition. Fig. 1 shows that during PVD, ultra-thin films roughen, with morphology of the same height scale as the film thickness. The film morphologies at thicknesses below 12~nm resemble semi-continuous morphologies typically observed in spinodal dewetting\cite{Xie1998,Farahzadi2010}. A uniform layer starts forming at thicknesses above 20~nm. For films thicker than 30~nm, the morphology flattens with both time and film thickness. This evolution in the morphology implies that during the deposition there is significant reconfiguration and motion of the molecules due to interfacial interactions, which allow for the formation of semi-continuous features. As the film thickness is increased, surface diffusion becomes more prominent and interfacial tension acts to smoothen the film. 

Since these films exhibit a spinodal morphology before a complete film is ever formed, the deposition rate and the characteristic length scale of the features can be used to estimate the average diffusion coefficient during PVD. Based on the spectral distribution shown in Fig. S4, in an 8~nm film this length scale is 350~nm. Given the deposition rate of 0.02~nm/s, it takes 600 seconds to deposit this film. Thus, the average diffusion coefficient is of the order of $3\times10^{-16}$~ m$^2$/s. For comparison, the bulk diffusion coefficient for most organic molecules at T$_{\mathrm{g}}$ is about $10^{-20}$~m$^2$/s\cite{Zhang2015}. This simple estimation implies that the average dynamics in 8 nm films, are several orders of magnitude faster than the bulk dynamics if measured at T$_{\mathrm{g}}$.

\begin{figure}[b]
	\includegraphics[width=.38\textwidth]{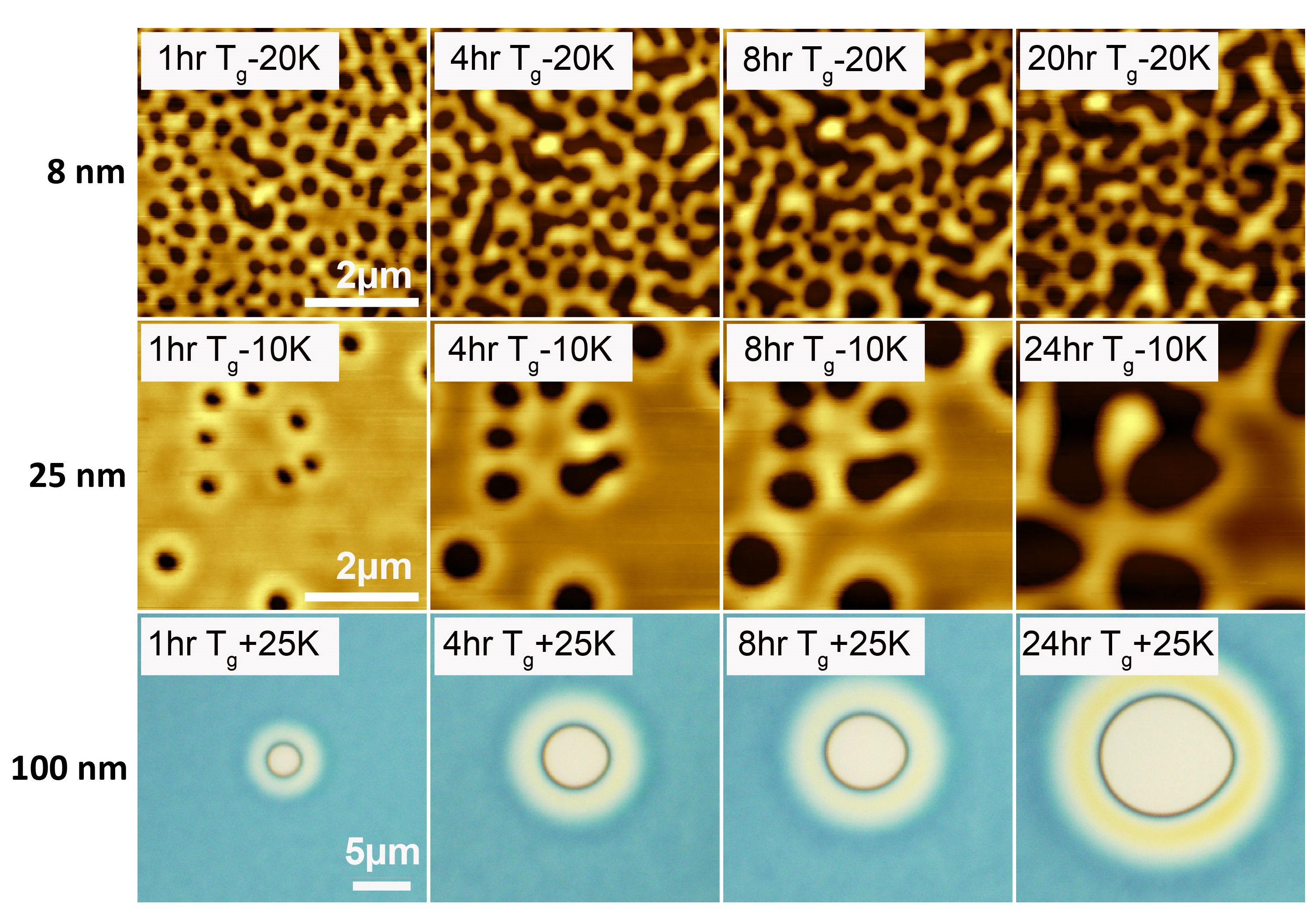}
	\caption{Time evolution of the morphology of TPD films during isothermal dewetting. All films were deposited at T$_{\mathrm{g}}$ with a deposition rate of 0.02~nm/s. Top row: 8~nm film held at T$_{\mathrm{g}}$-20~K. Middle row: 25~nm film held at T$_{\mathrm{g}}$-10~K; Bottom row: 100~nm film held at T$_{\mathrm{g}}$+25~K. Dewetting of 8~nm and 25~nm films were monitored by AFM, while 100~nm films were monitored by optical microscopy.}
\end{figure} 

The rough structures of as-deposited films were used as templates for further isothermal dewetting experiments. AFM or optical microscopy (OM), showed that the film morphology continued to evolve with time (examples shown in Fig. 2 and Fig. S5). Dewetting of ultra-thin films ($h<$~30~nm) progressed both through the growth of existing holes and bi-continuous features, as well as the nucleation and growth of new holes due to thermal capillary fluctuations. Isothermal dewetting was observed at temperatures as low as T$_{\mathrm{g}}$~-~30~K, where the bulk viscosity is not measurable and any reasonable extrapolation of the values of viscosity would predict a dewetting time longer than the age of the universe for a bulk film. As the holes continued to grow, material from the holes accumulated in rims. resulting in an increase in the film thickness outside the holes, which eventually stopped the process. In contrast, 100~nm films only dewetted well above T$_{\mathrm{g}}$ (T$>$T$_{\mathrm{g}}$~+~20~K), where the bulk viscosity is orders of magnitude lower. 

Due to the strong apparent film thickness-dependence of the dynamics, and non-uniform initial film morphologies, it is not possible to use models based on uniform viscosity and uniform film thickness\cite{Oron1997} to model the kinetics of dewetting in these films. Furthermore, the preexisting morphology can make the dewetting process appear faster, as the growth of existing holes in thin films is typically faster than the spontaneous nucleation of new holes in thick films (more details in SI). 
The substrate interaction potentials in these models\cite{Xie1998,Oron1997} are also poorly understood and slip condition at substrate interface is not explicitly included. 

\begin{figure}[b]
	\includegraphics[width=.43\textwidth]{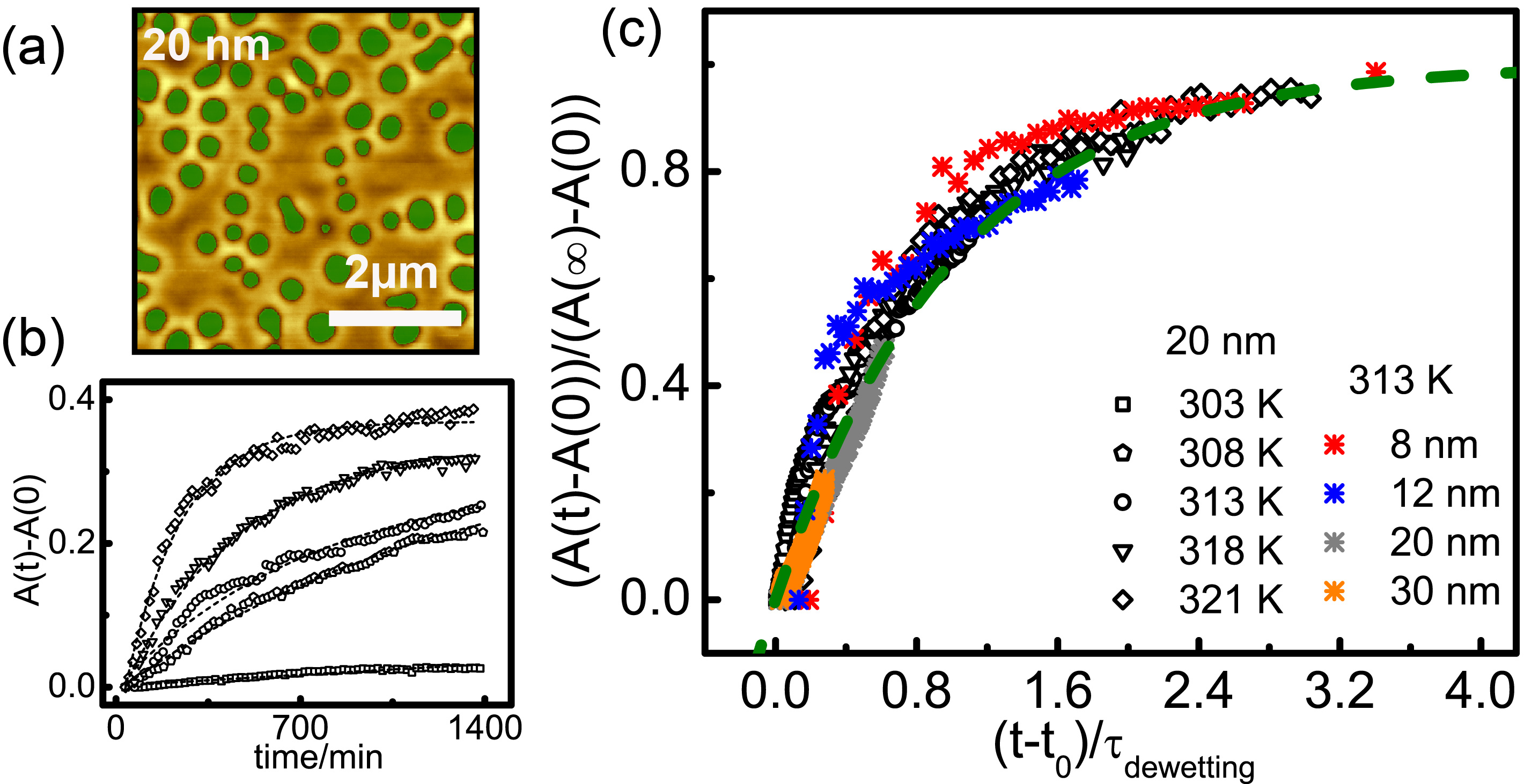}
	\caption{Dewetting kinetics tracked by AFM. (a) An appropriate height threshold is set to obtain the total dewetted area, marked as green. (b) Dewetted area, $A(t)$, as a function of time for 20~nm films at five annealing temperatures. (c) h-t-T superposition, relative dewetted area as a function of reduced time $t/\tau_{\mathrm{dewetting}}$. Black, open symbols are data from 20~nm films, shown in 2(b). Colored asterisks show dewetting of 8~nm, 12~nm, 20~nm, and 30~nm films at 313~K. Dashed green line is the universal fit function $\mathrm{y} = 1 - \mathrm{exp}(- \mathrm{t} / \textbf{{$\tau $}})$. Each data point in (b) and (c) is obtained from a single AFM image.}
\end{figure}

Despite these difficulties the effective viscosity can be indirectly measured by investigating the temperature dependence of the characteristic dewetting time, $\tau_{\mathrm{dewetting}}$\cite{Meredith2000}. This is because substrate interactions, surface tension, and the film's initial morphology are all weak functions of temperature, leaving the film's effective viscosity as the only temperature-dependent parameter driving dewetting (more details in SI). As such, $\tau_{\mathrm{dewetting}}$ should be proportional to the effective viscosity of the film, and thus the average structural relaxation time, $\tau_{\alpha}$. $\tau_{\mathrm{dewetting}}$, was measured by tracking the time evolution of the total dewetted area, $A(t)$, indicated by green color in Fig. 3(a) and fitting to a single exponential function \cite{Meredith2000}. Fig. 3(b) shows the change in $A(t)$ as a function of time for 20~nm films at various temperatures. Fig. 3(c) shows the normalized total dewetted area as a function of reduced time, $t/\tau_{\mathrm{dewetting}}$, for data at various annealing temperatures and film thicknesses. This figure shows that data for all films at a common thickness can be characterized with a single $\tau_{\mathrm{dewetting}}$. 


  If dynamics of thin films were identical to those of bulk, one would expect the temperature dependence of $\tau_{\mathrm{dewetting}}$ to be the same regardless of film thickness, even if absolute dewetting times depend on the film thickness due to different initial morphologies, and non-trivial thickness-dependence of the driving forces of dewetting. Fig. 4(a) shows an Arrhenius plot of -$\log\left(\tau_{\mathrm{dewetting}}\right)$ versus inverse temperature, $\mathrm{1/T}$. It is evident that in this temperature range, the slopes of the curves, which represent the apparent thermal activation barriers for rearrangement, E$_{\mathrm{a}}$ have  strong thickness dependence. For an 8~nm film $\tau_{\mathrm{dewetting}}$, and therefore the effective viscosity, changes less than half a decade over the temperature range of 298~K - 321~K. $\tau_{\mathrm{dewetting}}$ for 30~nm films show a much stronger temperature dependence, changing more than two decades over the same temperature range. It is important to note that all of these temperatures are well below bulk T$_{\mathrm{g}}$. The low apparent activation energy of ultra-thin films is consistent with previous studies of dynamics on polymeric thin films\cite{Yang2010,Glor2014,Glor2015}.

\begin{figure}[b]
	\includegraphics[width=.47\textwidth]{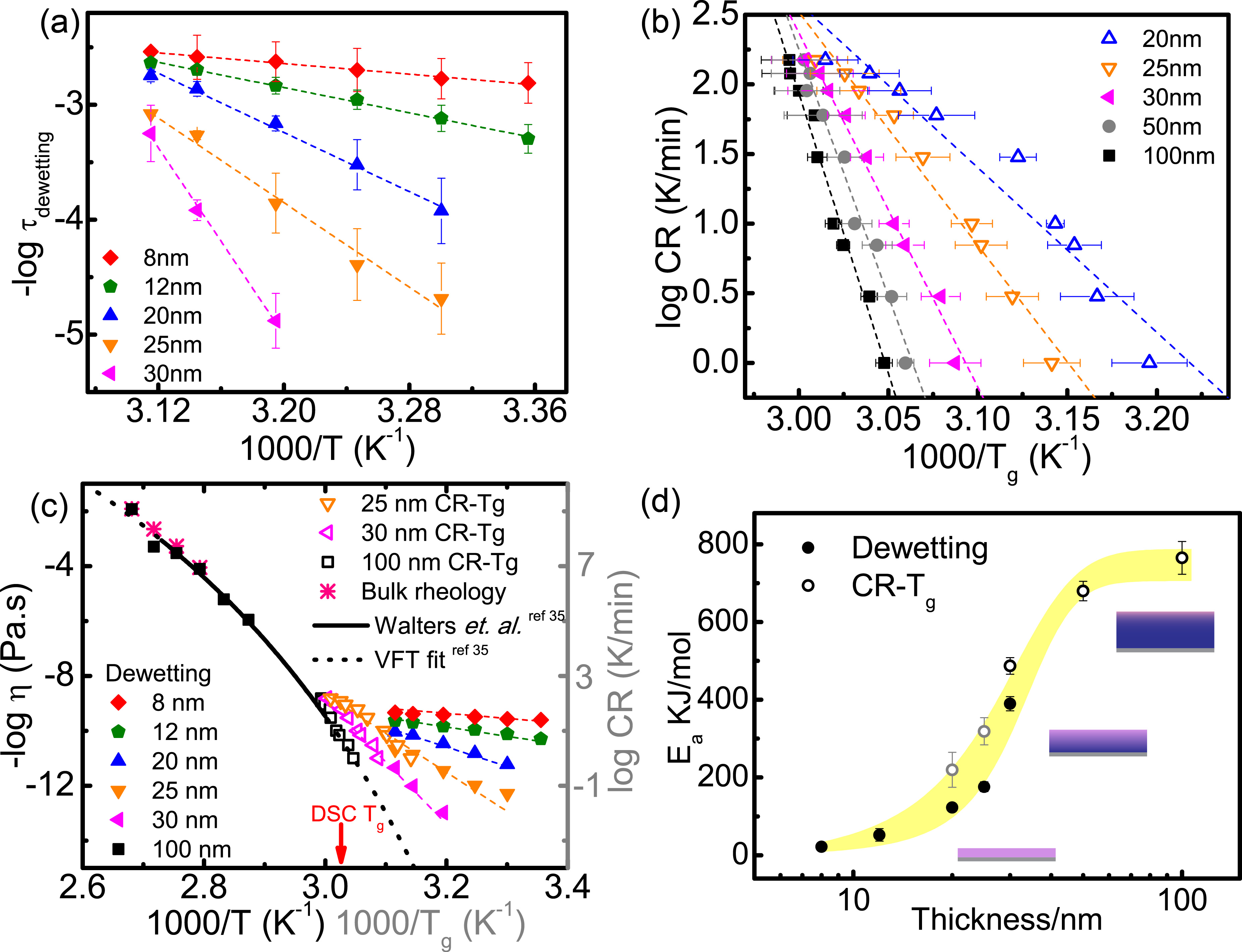}
	\caption{(a) $\tau_{\mathrm{dewetting}}$ vs. 1/T for various thicknesses. Dashed lines are Arrhenius fits. (b) Cooling rate vs. T$_{\mathrm{g}}$ for films of various thicknesses. Data marked with open symbols for 20~nm and 25~nm T$_{\mathrm{g}}$ values indicate potential uncertainties due to onset of dewetting and broadening of T$_{\mathrm{g}}$. (c) Viscosity (left axis) vs. 1/T obtained by rheology (pink asterisk) compared with dewetting data from (a) (filled symbols), and CR-T$_{\mathrm{g}}$ (right axis) from (b) (open symbols). Solid line represents dielectric relaxation data by Walter's {\it et al.}\cite{Walters2015}. Dotted black line is VFT fit to bulk viscosity. (d) Apparent activation energy, E$_\mathrm{a}$ obtained from Arrhenius fits of 4(a) (filled) and (b) (open) as a function of film thickness. Shaded area is a guide to the eyes for potential values of activation energy as measured by various methods.}
\end{figure}

In order to relate $\tau_{\mathrm{dewetting}}$ to viscosity, other experiments are required to define the vertical shift factors. Rheology measurements of bulk TPD are shown in Fig. 4(c) (details in SI). CR-T$_{\mathrm{g}}$ measurements were used to extend the dynamical range of the bulk measurements to temperatures below bulk T$_{\mathrm{g}}$\cite{Glor2015,Glor2014,Liu2015}. CR-T$_{\mathrm{g}}$ measurements were performed on films with a thickness range of 20~nm$<h<$100~nm as shown in Fig 4(b) (details in SI Fig. S8 and S9).  

Fig. 4(c) shows the combined data of dewetting and CR-T$_{\mathrm{g}}$ measurements, and provides a direct comparison between the bulk and the effective thin film viscosities. The dewetting measurements for 100~nm films were vertically shifted to match bulk rheology and dielectric relaxation measurements\cite{Walters2015} (Fig. S12). The CR-T$_{\mathrm{g}}$ data for 100~nm and 30~nm films were also shifted according to the relationship detailed in SI, $\log\eta=-\log(\mathrm{CR})+11$. In the temperature range of these experiments, there is excellent agreement between these three methods in determining viscosity. This strongly indicates that firstly, 100~nm films behave bulk-like and their effective viscosity matches that of bulk viscosity in the entire dynamical range of these measurements for both types of experiments, and secondly, other related parameters for the dewetting process, such as substrate interaction energy and surface tension, did not have strong temperature dependences. 

As detailed above, the initial morphology of films of various thicknesses are different. As such, the shift factor used to match the 100~nm dewetting data to bulk viscosity is not applicable to other films. Instead, for films of 25-30~nm, the CR-T$_{\mathrm{g}}$ experiments were used to calculate the appropriate shift factors (Fig. S12). 
It is important to note that CR-T$_{\mathrm{g}}$ could not be reliably used for ultra-thin films ($h\leq20$~nm) as detailed in SI. As such, the exact shift factors for these data sets are unknown and the data presented here for these films only reflect the temperature dependence of the effective viscosity and not their exact values. However, based on the simple analysis of diffusion presented earlier, the effective viscosity is at least about a factor of four faster than that of bulk at T$_{\mathrm{g}}$.

Fig. 4(c) provides a clear picture of the extent by which the dynamics are enhanced in ultra-thin films. While at bulk T$_{\mathrm{g}}$ the dynamics are enhanced by only one or two orders of magnitude, the difference between thin film and bulk dynamics continues to diverge as the temperature is decreased below T$_{\mathrm{g}}$. For example, at a temperature of T$_{\mathrm{g}}$-35~K, the bulk viscosity becomes unmeasurable, while the effective thin film viscosity only changes by less than two orders of magnitude from the value at T$_{\mathrm{g}}$. Fig. 4(d) shows the apparent activation energy, E$_\mathrm{a}$ (slope of log viscosity vs. $\mathrm{1/T}$) as a function of film thickness as determined via both dewetting and CR-T$_{\mathrm{g}}$ experiments, with dewetting experiments setting the lower bound and CR  T$_{\mathrm{g}}$ setting the upper bound value for E$_\mathrm{a}$ (Details discussed in SI). We note that the temperature range of these experiments are limited and E$_\mathrm{a}$ may vary with temperature closer to or above T$_{\mathrm{g}}$. 

As shown in Fig. 4(d), in films with $h<$~20~nm, the E$_\mathrm{a}$ is much lower than that of bulk, and has a weak thickness dependence. The low activation barrier for dewetting confirms that the rough morphologies observed in ultra-thin PVD films are due to fast dewetting during deposition. E$_\mathrm{a}$ increases sharply in films with thicknesses between 20~nm$<h<$30~nm, and becomes very similar to bulk at $h>$~40~nm. In this regime, the dynamics of the film are bulk-like during PVD, and surface diffusion acts to smoothen the film. Interestingly, in ultra-thin films, once the local film thickness around the rims reaches 40~nm the dewetting process also slows down significantly and appears to stop (Fig. S7). This remarkably sharp transition in the dynamics suggests that the gradient of dynamics induced by the interfacial effects is not the same in films of different thicknesses, as schematically shown in the inset of Fig. 4(d). In ultra-thin films, the dynamics are enhanced in the entire film, showing little thickness dependence, while in films with thicknesses of 40~nm or more, the dynamics in the entire film is bulk-like except for perhaps a few liquid-like mono-layers near the free surface. It is important to note that the temperature dependence of diffusion coefficients measured on the surface of bulk films\cite{Zhu2011,Daley2012} are also larger than those measured in ultra-thin films in this study, further confirming that the dynamics at the surface of a bulk film are different than those measured in thin films. These observations suggest that the dynamics of the glassy material are correlated over large length scales and the dynamics of thin films are influenced both by the interfacial dynamics and the glassy dynamics in the layers closer to the center of the film. 

As such, surface diffusion measurements on the surface of bulk-like films alone are inadequate in predicting the activation barrier for the dynamics in ultra-thin films and the length scale of the effects. Direct measurements of properties as a function of film thickness are required for determining the correlation length for the dynamics. While to our knowledge there are no such prior studies in thin films of other organic glasses, observations in polymeric glasses\cite{Glor2015} show a similar non-linear transition in E$_\mathrm{a}$ as a function of film thickness, with the midpoint of transition in the 20-30~nm film thickness region. Earlier studies by Ellison and Torkelson also suggest correlated dynamics in the top and bottom layers of a polymeric film as the film thickness is reduced below 20~nm\cite{Ellison2003}. However, the transition appears to be much sharper in organic thin films with much lower activation barriers in ultra-thin film regime. This may imply that chain effects are also important in facilitating the dynamics in polymeric thin films. Regardless, the general similar trend and length scale of enhanced dynamics in both small organic molecular and polymeric glasses suggest that long range facilitation of the dynamics may be a characteristic feature of glassy systems. Future studies on more glassy systems are needed to confirm whether these observations are ubiquitous in organic glassy systems. Models based on a constant length scale of interfacial effects\cite{Hanakata2014,Stevenson2008} may not fully capture  the  strong,  almost sigmoidal transition in apparent activation barriers observed here. Instead, models based on long range elastic response\cite{Mirigian2014}, or which use growing cooperative length scales\cite{Hocky2012,Butler1991} may be able to predict such strong correlated dynamics.

In summary, we have presented that the temperature dependence of the effective viscosity, and thus the structural relaxation time, of thick and ultra-thin films of molecular glasses can be measured via a combination of isothermal dewetting and CR-T$_{\mathrm{g}}$ measurements. We have demonstrated that the rough initial morphology of vapor-deposited thin films are closely related to the enhanced dynamics in ultra-thin films. Films as thick as 30~nm dewet spontaneously well below bulk T$_{\mathrm{g}}$, indicative of greatly enhanced dynamics in these films. An examination of the thickness dependence of the apparent activation barrier in these films reveals a sharp, sigmoidal transition in the dynamics as the thickness varies between 20 to 40 nm indicating a strong correlation between the dynamics of the free surface and the bulk of the film. This implies an interplay between the facilitation of the dynamics by the interface and the bulk glass, with a considerably large length scale of about 30~nm. 

\begin{acknowledgments}
We thank Dr. Wei-Shao Tung, Professor Karen Winey for help with rheology measurement, M. Reza Rahimi Tabar, Andrea Liu, Mark Ediger, Ramalingam Kailasham, Amit Shavit and Jack Douglas for helpful discussions. Z.F. and R.R. acknowledge seed funding from MRSEC grant (NSF DMR-1120901). K.W. was an undergraduate student enrolled in the Research Experiences for Undergraduates (REU) program during 2014 summer. He is from the Department of Physics, University of Texas Rio Grande Valley. He acknowledges funding from an NSF MRSEC grant (DMR-1120901).
\end{acknowledgments}

\bibliography{dewettingreferenceabbr}

\begin{thebibliography}{38}%
\makeatletter
\providecommand \@ifxundefined [1]{%
 \@ifx{#1\undefined}
}%
\providecommand \@ifnum [1]{%
 \ifnum #1\expandafter \@firstoftwo
 \else \expandafter \@secondoftwo
 \fi
}%
\providecommand \@ifx [1]{%
 \ifx #1\expandafter \@firstoftwo
 \else \expandafter \@secondoftwo
 \fi
}%
\providecommand \natexlab [1]{#1}%
\providecommand \enquote  [1]{``#1''}%
\providecommand \bibnamefont  [1]{#1}%
\providecommand \bibfnamefont [1]{#1}%
\providecommand \citenamefont [1]{#1}%
\providecommand \href@noop [0]{\@secondoftwo}%
\providecommand \href [0]{\begingroup \@sanitize@url \@href}%
\providecommand \@href[1]{\@@startlink{#1}\@@href}%
\providecommand \@@href[1]{\endgroup#1\@@endlink}%
\providecommand \@sanitize@url [0]{\catcode `\\12\catcode `\$12\catcode
  `\&12\catcode `\#12\catcode `\^12\catcode `\_12\catcode `\%12\relax}%
\providecommand \@@startlink[1]{}%
\providecommand \@@endlink[0]{}%
\providecommand \url  [0]{\begingroup\@sanitize@url \@url }%
\providecommand \@url [1]{\endgroup\@href {#1}{\urlprefix }}%
\providecommand \urlprefix  [0]{URL }%
\providecommand \Eprint [0]{\href }%
\providecommand \doibase [0]{http://dx.doi.org/}%
\providecommand \selectlanguage [0]{\@gobble}%
\providecommand \bibinfo  [0]{\@secondoftwo}%
\providecommand \bibfield  [0]{\@secondoftwo}%
\providecommand \translation [1]{[#1]}%
\providecommand \BibitemOpen [0]{}%
\providecommand \bibitemStop [0]{}%
\providecommand \bibitemNoStop [0]{.\EOS\space}%
\providecommand \EOS [0]{\spacefactor3000\relax}%
\providecommand \BibitemShut  [1]{\csname bibitem#1\endcsname}%
\let\auto@bib@innerbib\@empty
\bibitem [{\citenamefont {Shirota}\ and\ \citenamefont
  {Kageyama}(2007)}]{Shirota2007}%
  \BibitemOpen
  \bibfield  {author} {\bibinfo {author} {\bibfnamefont {Y.}~\bibnamefont
  {Shirota}}\ and\ \bibinfo {author} {\bibfnamefont {H.}~\bibnamefont
  {Kageyama}},\ }\href {\doibase 10.1021/cr050143} {\bibfield  {journal}
  {\bibinfo  {journal} {Chem. Rev.}\ }\textbf {\bibinfo {volume} {107}},\
  \bibinfo {pages} {953} (\bibinfo {year} {2007})}\BibitemShut {NoStop}%
\bibitem [{\citenamefont {Veinot}\ and\ \citenamefont
  {Marks}(2005)}]{Veinot2005}%
  \BibitemOpen
  \bibfield  {author} {\bibinfo {author} {\bibfnamefont {J.~G.~C.}\
  \bibnamefont {Veinot}}\ and\ \bibinfo {author} {\bibfnamefont {T.~J.}\
  \bibnamefont {Marks}},\ }\href {\doibase 10.1021/ar030210r} {\bibfield
  {journal} {\bibinfo  {journal} {Acc. Chem. Res.}\ }\textbf {\bibinfo {volume}
  {38}},\ \bibinfo {pages} {632} (\bibinfo {year} {2005})}\BibitemShut
  {NoStop}%
\bibitem [{\citenamefont {Tang}\ and\ \citenamefont
  {Vanslyke}(1987)}]{Tang1987}%
  \BibitemOpen
  \bibfield  {author} {\bibinfo {author} {\bibfnamefont {C.~W.}\ \bibnamefont
  {Tang}}\ and\ \bibinfo {author} {\bibfnamefont {S.~A.}\ \bibnamefont
  {Vanslyke}},\ }\href {\doibase 10.1063/1.98799} {\bibfield  {journal}
  {\bibinfo  {journal} {Appl. Phys. Lett.}\ }\textbf {\bibinfo {volume} {51}},\
  \bibinfo {pages} {913} (\bibinfo {year} {1987})}\BibitemShut {NoStop}%
\bibitem [{\citenamefont {Mahltig}\ \emph {et~al.}(2005)\citenamefont
  {Mahltig}, \citenamefont {B\"{o}ttcher}, \citenamefont {Rauch}, \citenamefont
  {Dieckmann}, \citenamefont {Nitsche},\ and\ \citenamefont
  {Fritz}}]{Mahltig2005}%
  \BibitemOpen
  \bibfield  {author} {\bibinfo {author} {\bibfnamefont {B.}~\bibnamefont
  {Mahltig}}, \bibinfo {author} {\bibfnamefont {H.}~\bibnamefont
  {B\"{o}ttcher}}, \bibinfo {author} {\bibfnamefont {K.}~\bibnamefont {Rauch}},
  \bibinfo {author} {\bibfnamefont {U.}~\bibnamefont {Dieckmann}}, \bibinfo
  {author} {\bibfnamefont {R.}~\bibnamefont {Nitsche}}, \ and\ \bibinfo
  {author} {\bibfnamefont {T.}~\bibnamefont {Fritz}},\ }\href {\doibase
  10.1016/j.tsf.2005.03.056} {\bibfield  {journal} {\bibinfo  {journal} {Thin
  Solid Films}\ }\textbf {\bibinfo {volume} {485}},\ \bibinfo {pages} {108}
  (\bibinfo {year} {2005})}\BibitemShut {NoStop}%
\bibitem [{\citenamefont {Pires}\ \emph {et~al.}(2010)\citenamefont {Pires},
  \citenamefont {Hedrick}, \citenamefont {{De Silva}}, \citenamefont {Frommer},
  \citenamefont {Gotsmann}, \citenamefont {Wolf}, \citenamefont {Despont},
  \citenamefont {Duerig},\ and\ \citenamefont {Knoll}}]{Pires2010}%
  \BibitemOpen
  \bibfield  {author} {\bibinfo {author} {\bibfnamefont {D.}~\bibnamefont
  {Pires}}, \bibinfo {author} {\bibfnamefont {J.~L.}\ \bibnamefont {Hedrick}},
  \bibinfo {author} {\bibfnamefont {A.}~\bibnamefont {{De Silva}}}, \bibinfo
  {author} {\bibfnamefont {J.}~\bibnamefont {Frommer}}, \bibinfo {author}
  {\bibfnamefont {B.}~\bibnamefont {Gotsmann}}, \bibinfo {author}
  {\bibfnamefont {H.}~\bibnamefont {Wolf}}, \bibinfo {author} {\bibfnamefont
  {M.}~\bibnamefont {Despont}}, \bibinfo {author} {\bibfnamefont
  {U.}~\bibnamefont {Duerig}}, \ and\ \bibinfo {author} {\bibfnamefont {A.~W.}\
  \bibnamefont {Knoll}},\ }\href {\doibase 10.1126/science.1187851} {\bibfield
  {journal} {\bibinfo  {journal} {Science}\ }\textbf {\bibinfo {volume}
  {328}},\ \bibinfo {pages} {732} (\bibinfo {year} {2010})}\BibitemShut
  {NoStop}%
\bibitem [{\citenamefont {Zhu}\ \emph {et~al.}(2011)\citenamefont {Zhu},
  \citenamefont {Brian}, \citenamefont {Swallen}, \citenamefont {Straus},
  \citenamefont {Ediger},\ and\ \citenamefont {Yu}}]{Zhu2011}%
  \BibitemOpen
  \bibfield  {author} {\bibinfo {author} {\bibfnamefont {L.}~\bibnamefont
  {Zhu}}, \bibinfo {author} {\bibfnamefont {C.~W.}\ \bibnamefont {Brian}},
  \bibinfo {author} {\bibfnamefont {S.~F.}\ \bibnamefont {Swallen}}, \bibinfo
  {author} {\bibfnamefont {P.~T.}\ \bibnamefont {Straus}}, \bibinfo {author}
  {\bibfnamefont {M.~D.}\ \bibnamefont {Ediger}}, \ and\ \bibinfo {author}
  {\bibfnamefont {L.}~\bibnamefont {Yu}},\ }\href@noop {} {\bibfield  {journal}
  {\bibinfo  {journal} {Phys. Rev. Lett.}\ }\textbf {\bibinfo {volume} {106}},\
  \bibinfo {pages} {256103} (\bibinfo {year} {2011})}\BibitemShut {NoStop}%
\bibitem [{\citenamefont {Daley}\ \emph {et~al.}(2012)\citenamefont {Daley},
  \citenamefont {Fakhraai}, \citenamefont {Ediger},\ and\ \citenamefont
  {Forrest}}]{Daley2012}%
  \BibitemOpen
  \bibfield  {author} {\bibinfo {author} {\bibfnamefont {C.~R.}\ \bibnamefont
  {Daley}}, \bibinfo {author} {\bibfnamefont {Z.}~\bibnamefont {Fakhraai}},
  \bibinfo {author} {\bibfnamefont {M.~D.}\ \bibnamefont {Ediger}}, \ and\
  \bibinfo {author} {\bibfnamefont {J.~A.}\ \bibnamefont {Forrest}},\ }\href
  {\doibase 10.1039/c2sm06826e} {\bibfield  {journal} {\bibinfo  {journal}
  {Soft Matter}\ }\textbf {\bibinfo {volume} {8}},\ \bibinfo {pages} {2206}
  (\bibinfo {year} {2012})}\BibitemShut {NoStop}%
\bibitem [{\citenamefont {Fakhraai}\ and\ \citenamefont
  {Forrest}(2008)}]{Fakhraai2008}%
  \BibitemOpen
  \bibfield  {author} {\bibinfo {author} {\bibfnamefont {Z.}~\bibnamefont
  {Fakhraai}}\ and\ \bibinfo {author} {\bibfnamefont {J.~A.}\ \bibnamefont
  {Forrest}},\ }\href {\doibase 10.1126/science.1151205} {\bibfield  {journal}
  {\bibinfo  {journal} {Science}\ }\textbf {\bibinfo {volume} {319}},\ \bibinfo
  {pages} {600} (\bibinfo {year} {2008})}\BibitemShut {NoStop}%
\bibitem [{\citenamefont {Paeng}\ \emph {et~al.}(2011)\citenamefont {Paeng},
  \citenamefont {Swallen},\ and\ \citenamefont {Ediger}}]{Paeng2011}%
  \BibitemOpen
  \bibfield  {author} {\bibinfo {author} {\bibfnamefont {K.}~\bibnamefont
  {Paeng}}, \bibinfo {author} {\bibfnamefont {S.~F.}\ \bibnamefont {Swallen}},
  \ and\ \bibinfo {author} {\bibfnamefont {M.~D.}\ \bibnamefont {Ediger}},\
  }\href {\doibase 10.1021/ja2022834} {\bibfield  {journal} {\bibinfo
  {journal} {J. Am. Chem. Soc.}\ }\textbf {\bibinfo {volume} {133}},\ \bibinfo
  {pages} {8444} (\bibinfo {year} {2011})}\BibitemShut {NoStop}%
\bibitem [{\citenamefont {Keddie}\ \emph {et~al.}(1994)\citenamefont {Keddie},
  \citenamefont {Jones},\ and\ \citenamefont {Corey}}]{Keddie1994}%
  \BibitemOpen
  \bibfield  {author} {\bibinfo {author} {\bibfnamefont {J.~L.}\ \bibnamefont
  {Keddie}}, \bibinfo {author} {\bibfnamefont {R.~A.~L.}\ \bibnamefont
  {Jones}}, \ and\ \bibinfo {author} {\bibfnamefont {R.~A.}\ \bibnamefont
  {Corey}},\ }\href@noop {} {\bibfield  {journal} {\bibinfo  {journal}
  {Europhys. Lett.}\ }\textbf {\bibinfo {volume} {27}},\ \bibinfo {pages} {59}
  (\bibinfo {year} {1994})}\BibitemShut {NoStop}%
\bibitem [{\citenamefont {Ediger}\ and\ \citenamefont
  {Forrest}(2014)}]{Ediger2014}%
  \BibitemOpen
  \bibfield  {author} {\bibinfo {author} {\bibfnamefont {M.~D.}\ \bibnamefont
  {Ediger}}\ and\ \bibinfo {author} {\bibfnamefont {J.~A.}\ \bibnamefont
  {Forrest}},\ }\href {\doibase 10.1021/ma4017696} {\bibfield  {journal}
  {\bibinfo  {journal} {Macromolecules}\ }\textbf {\bibinfo {volume} {47}},\
  \bibinfo {pages} {471} (\bibinfo {year} {2014})}\BibitemShut {NoStop}%
\bibitem [{\citenamefont {Ellison}\ and\ \citenamefont
  {Torkelson}(2003)}]{Ellison2003}%
  \BibitemOpen
  \bibfield  {author} {\bibinfo {author} {\bibfnamefont {C.~J.}\ \bibnamefont
  {Ellison}}\ and\ \bibinfo {author} {\bibfnamefont {J.~M.}\ \bibnamefont
  {Torkelson}},\ }\href {\doibase 10.1038/nmat980} {\bibfield  {journal}
  {\bibinfo  {journal} {Nat. Mater.}\ }\textbf {\bibinfo {volume} {2}},\
  \bibinfo {pages} {695} (\bibinfo {year} {2003})}\BibitemShut {NoStop}%
\bibitem [{\citenamefont {Shavit}\ and\ \citenamefont
  {Riggleman}(2013)}]{Shavit2013}%
  \BibitemOpen
  \bibfield  {author} {\bibinfo {author} {\bibfnamefont {A.}~\bibnamefont
  {Shavit}}\ and\ \bibinfo {author} {\bibfnamefont {R.~A.}\ \bibnamefont
  {Riggleman}},\ }\href {\doibase 10.1021/ma400210w} {\bibfield  {journal}
  {\bibinfo  {journal} {Macromolecules}\ }\textbf {\bibinfo {volume} {46}},\
  \bibinfo {pages} {5044} (\bibinfo {year} {2013})}\BibitemShut {NoStop}%
\bibitem [{\citenamefont {Glor}\ and\ \citenamefont
  {Fakhraai}(2014)}]{Glor2014}%
  \BibitemOpen
  \bibfield  {author} {\bibinfo {author} {\bibfnamefont {E.~C.}\ \bibnamefont
  {Glor}}\ and\ \bibinfo {author} {\bibfnamefont {Z.}~\bibnamefont
  {Fakhraai}},\ }\href {\doibase 10.1063/1.4901512} {\bibfield  {journal}
  {\bibinfo  {journal} {J. Chem. Phys.}\ }\textbf {\bibinfo {volume} {141}},\
  \bibinfo {pages} {194505} (\bibinfo {year} {2014})}\BibitemShut {NoStop}%
\bibitem [{\citenamefont {Yang}\ \emph {et~al.}(2010)\citenamefont {Yang},
  \citenamefont {Fujii}, \citenamefont {Lee}, \citenamefont {Lam},\ and\
  \citenamefont {Tsui}}]{Yang2010}%
  \BibitemOpen
  \bibfield  {author} {\bibinfo {author} {\bibfnamefont {Z.}~\bibnamefont
  {Yang}}, \bibinfo {author} {\bibfnamefont {Y.}~\bibnamefont {Fujii}},
  \bibinfo {author} {\bibfnamefont {F.~K.}\ \bibnamefont {Lee}}, \bibinfo
  {author} {\bibfnamefont {C.-H.}\ \bibnamefont {Lam}}, \ and\ \bibinfo
  {author} {\bibfnamefont {O.~K.~C.}\ \bibnamefont {Tsui}},\ }\href@noop {}
  {\bibfield  {journal} {\bibinfo  {journal} {Science}\ }\textbf {\bibinfo
  {volume} {328}},\ \bibinfo {pages} {1676} (\bibinfo {year}
  {2010})}\BibitemShut {NoStop}%
\bibitem [{\citenamefont {Pye}\ \emph {et~al.}(2010)\citenamefont {Pye},
  \citenamefont {Rohald}, \citenamefont {Baker},\ and\ \citenamefont
  {Roth}}]{Pye2010}%
  \BibitemOpen
  \bibfield  {author} {\bibinfo {author} {\bibfnamefont {J.~E.}\ \bibnamefont
  {Pye}}, \bibinfo {author} {\bibfnamefont {K.~A.}\ \bibnamefont {Rohald}},
  \bibinfo {author} {\bibfnamefont {E.~A.}\ \bibnamefont {Baker}}, \ and\
  \bibinfo {author} {\bibfnamefont {C.~B.}\ \bibnamefont {Roth}},\ }\href
  {\doibase 10.1021/ma101412r} {\bibfield  {journal} {\bibinfo  {journal}
  {Macromolecules}\ }\textbf {\bibinfo {volume} {43}},\ \bibinfo {pages} {8296}
  (\bibinfo {year} {2010})}\BibitemShut {NoStop}%
\bibitem [{\citenamefont {Priestley}\ \emph {et~al.}(2005)\citenamefont
  {Priestley}, \citenamefont {Ellison}, \citenamefont {Broadbelt},\ and\
  \citenamefont {Torkelson}}]{Priestley2005}%
  \BibitemOpen
  \bibfield  {author} {\bibinfo {author} {\bibfnamefont {R.~D.}\ \bibnamefont
  {Priestley}}, \bibinfo {author} {\bibfnamefont {C.~J.}\ \bibnamefont
  {Ellison}}, \bibinfo {author} {\bibfnamefont {L.~J.}\ \bibnamefont
  {Broadbelt}}, \ and\ \bibinfo {author} {\bibfnamefont {J.~M.}\ \bibnamefont
  {Torkelson}},\ }\href {\doibase 10.1126/science.1112217} {\bibfield
  {journal} {\bibinfo  {journal} {Science}\ }\textbf {\bibinfo {volume}
  {309}},\ \bibinfo {pages} {456} (\bibinfo {year} {2005})}\BibitemShut
  {NoStop}%
\bibitem [{\citenamefont {Sep\'{u}lveda}\ \emph {et~al.}(2011)\citenamefont
  {Sep\'{u}lveda}, \citenamefont {Leon-Gutierrez}, \citenamefont
  {Gonzalez-Silveira}, \citenamefont {Rodr\'{\i}guez-Tinoco}, \citenamefont
  {Clavaguera-Mora},\ and\ \citenamefont
  {Rodr\'{\i}guez-Viejo}}]{Sepulveda2011}%
  \BibitemOpen
  \bibfield  {author} {\bibinfo {author} {\bibfnamefont {A.}~\bibnamefont
  {Sep\'{u}lveda}}, \bibinfo {author} {\bibfnamefont {E.}~\bibnamefont
  {Leon-Gutierrez}}, \bibinfo {author} {\bibfnamefont {M.}~\bibnamefont
  {Gonzalez-Silveira}}, \bibinfo {author} {\bibfnamefont {C.}~\bibnamefont
  {Rodr\'{\i}guez-Tinoco}}, \bibinfo {author} {\bibfnamefont {M.~T.}\
  \bibnamefont {Clavaguera-Mora}}, \ and\ \bibinfo {author} {\bibfnamefont
  {J.}~\bibnamefont {Rodr\'{\i}guez-Viejo}},\ }\href {\doibase
  10.1103/PhysRevLett.107.025901} {\bibfield  {journal} {\bibinfo  {journal}
  {Phys. Rev. Lett.}\ }\textbf {\bibinfo {volume} {107}},\ \bibinfo {pages}
  {025901} (\bibinfo {year} {2011})}\BibitemShut {NoStop}%
\bibitem [{\citenamefont {Swallen}\ \emph {et~al.}(2007)\citenamefont
  {Swallen}, \citenamefont {Kearns}, \citenamefont {Mapes}, \citenamefont
  {Kim}, \citenamefont {McMahon}, \citenamefont {Ediger}, \citenamefont {Wu},
  \citenamefont {Yu},\ and\ \citenamefont {Satija}}]{Swallen2007}%
  \BibitemOpen
  \bibfield  {author} {\bibinfo {author} {\bibfnamefont {S.~F.}\ \bibnamefont
  {Swallen}}, \bibinfo {author} {\bibfnamefont {K.~L.}\ \bibnamefont {Kearns}},
  \bibinfo {author} {\bibfnamefont {M.~K.}\ \bibnamefont {Mapes}}, \bibinfo
  {author} {\bibfnamefont {Y.~S.}\ \bibnamefont {Kim}}, \bibinfo {author}
  {\bibfnamefont {R.~J.}\ \bibnamefont {McMahon}}, \bibinfo {author}
  {\bibfnamefont {M.~D.}\ \bibnamefont {Ediger}}, \bibinfo {author}
  {\bibfnamefont {T.}~\bibnamefont {Wu}}, \bibinfo {author} {\bibfnamefont
  {L.}~\bibnamefont {Yu}}, \ and\ \bibinfo {author} {\bibfnamefont
  {S.}~\bibnamefont {Satija}},\ }\href {\doibase 10.1126/science.1135795}
  {\bibfield  {journal} {\bibinfo  {journal} {Science}\ }\textbf {\bibinfo
  {volume} {315}},\ \bibinfo {pages} {353} (\bibinfo {year}
  {2007})}\BibitemShut {NoStop}%
\bibitem [{\citenamefont {Leon-Gutierrez}\ \emph {et~al.}(2010)\citenamefont
  {Leon-Gutierrez}, \citenamefont {Sep\'{u}lveda}, \citenamefont {Garcia},
  \citenamefont {Clavaguera-Mora},\ and\ \citenamefont
  {Rodr\'{\i}guez-Viejo}}]{Leon-Gutierrez2010}%
  \BibitemOpen
  \bibfield  {author} {\bibinfo {author} {\bibfnamefont {E.}~\bibnamefont
  {Leon-Gutierrez}}, \bibinfo {author} {\bibfnamefont {A.}~\bibnamefont
  {Sep\'{u}lveda}}, \bibinfo {author} {\bibfnamefont {G.}~\bibnamefont
  {Garcia}}, \bibinfo {author} {\bibfnamefont {M.~T.}\ \bibnamefont
  {Clavaguera-Mora}}, \ and\ \bibinfo {author} {\bibfnamefont {J.}~\bibnamefont
  {Rodr\'{\i}guez-Viejo}},\ }\href {\doibase 10.1039/c0cp00208a} {\bibfield
  {journal} {\bibinfo  {journal} {Phys. Chem. Chem. Phys.}\ }\textbf {\bibinfo
  {volume} {12}},\ \bibinfo {pages} {14693} (\bibinfo {year}
  {2010})}\BibitemShut {NoStop}%
\bibitem [{\citenamefont {Fakhraai}\ \emph {et~al.}(2011)\citenamefont
  {Fakhraai}, \citenamefont {Still}, \citenamefont {Fytas},\ and\ \citenamefont
  {Ediger}}]{Fakhraai2011}%
  \BibitemOpen
  \bibfield  {author} {\bibinfo {author} {\bibfnamefont {Z.}~\bibnamefont
  {Fakhraai}}, \bibinfo {author} {\bibfnamefont {T.}~\bibnamefont {Still}},
  \bibinfo {author} {\bibfnamefont {G.}~\bibnamefont {Fytas}}, \ and\ \bibinfo
  {author} {\bibfnamefont {M.~D.}\ \bibnamefont {Ediger}},\ }\href {\doibase
  10.1021/jz101723d} {\bibfield  {journal} {\bibinfo  {journal} {J. Phys. Chem.
  Lett.}\ }\textbf {\bibinfo {volume} {2}},\ \bibinfo {pages} {423} (\bibinfo
  {year} {2011})}\BibitemShut {NoStop}%
\bibitem [{\citenamefont {Liu}\ \emph {et~al.}(2015{\natexlab{a}})\citenamefont
  {Liu}, \citenamefont {Cheng}, \citenamefont {Salami-Ranjbaran}, \citenamefont
  {Gao}, \citenamefont {Li}, \citenamefont {Tong}, \citenamefont {Lin},
  \citenamefont {Zhang}, \citenamefont {Zhang}, \citenamefont {Klinge},
  \citenamefont {Walsh},\ and\ \citenamefont {Fakhraai}}]{Liu2015a}%
  \BibitemOpen
  \bibfield  {author} {\bibinfo {author} {\bibfnamefont {T.}~\bibnamefont
  {Liu}}, \bibinfo {author} {\bibfnamefont {K.}~\bibnamefont {Cheng}}, \bibinfo
  {author} {\bibfnamefont {E.}~\bibnamefont {Salami-Ranjbaran}}, \bibinfo
  {author} {\bibfnamefont {F.}~\bibnamefont {Gao}}, \bibinfo {author}
  {\bibfnamefont {C.}~\bibnamefont {Li}}, \bibinfo {author} {\bibfnamefont
  {X.}~\bibnamefont {Tong}}, \bibinfo {author} {\bibfnamefont {Y.-C.}\
  \bibnamefont {Lin}}, \bibinfo {author} {\bibfnamefont {Y.}~\bibnamefont
  {Zhang}}, \bibinfo {author} {\bibfnamefont {W.}~\bibnamefont {Zhang}},
  \bibinfo {author} {\bibfnamefont {L.}~\bibnamefont {Klinge}}, \bibinfo
  {author} {\bibfnamefont {P.~J.}\ \bibnamefont {Walsh}}, \ and\ \bibinfo
  {author} {\bibfnamefont {Z.}~\bibnamefont {Fakhraai}},\ }\href {\doibase
  10.1063/1.4928521} {\bibfield  {journal} {\bibinfo  {journal} {J. Chem.
  Phys.}\ }\textbf {\bibinfo {volume} {143}},\ \bibinfo {pages} {084506}
  (\bibinfo {year} {2015}{\natexlab{a}})}\BibitemShut {NoStop}%
\bibitem [{\citenamefont {Chua}\ \emph {et~al.}(2015)\citenamefont {Chua},
  \citenamefont {Ahrenberg}, \citenamefont {Tylinski}, \citenamefont {Ediger},\
  and\ \citenamefont {Schick}}]{Chua2015}%
  \BibitemOpen
  \bibfield  {author} {\bibinfo {author} {\bibfnamefont {Y.~Z.}\ \bibnamefont
  {Chua}}, \bibinfo {author} {\bibfnamefont {M.}~\bibnamefont {Ahrenberg}},
  \bibinfo {author} {\bibfnamefont {M.}~\bibnamefont {Tylinski}}, \bibinfo
  {author} {\bibfnamefont {M.~D.}\ \bibnamefont {Ediger}}, \ and\ \bibinfo
  {author} {\bibfnamefont {C.}~\bibnamefont {Schick}},\ }\href {\doibase
  10.1063/1.4906806} {\bibfield  {journal} {\bibinfo  {journal} {J. Chem.
  Phys.}\ }\textbf {\bibinfo {volume} {142}},\ \bibinfo {pages} {054506}
  (\bibinfo {year} {2015})}\BibitemShut {NoStop}%
\bibitem [{\citenamefont {Singh}\ and\ \citenamefont {{De
  Pablo}}(2011)}]{Singh2011a}%
  \BibitemOpen
  \bibfield  {author} {\bibinfo {author} {\bibfnamefont {S.}~\bibnamefont
  {Singh}}\ and\ \bibinfo {author} {\bibfnamefont {J.~J.}\ \bibnamefont {{De
  Pablo}}},\ }\href {\doibase 10.1063/1.3586805} {\bibfield  {journal}
  {\bibinfo  {journal} {J. Chem. Phys.}\ }\textbf {\bibinfo {volume} {134}},\
  \bibinfo {pages} {194903} (\bibinfo {year} {2011})}\BibitemShut {NoStop}%
\bibitem [{\citenamefont {{Pazmi\~{n}o Betancourt}}\ \emph
  {et~al.}(2015)\citenamefont {{Pazmi\~{n}o Betancourt}}, \citenamefont
  {Hanakata}, \citenamefont {Starr},\ and\ \citenamefont
  {Douglas}}]{PazminoBetancourt2015}%
  \BibitemOpen
  \bibfield  {author} {\bibinfo {author} {\bibfnamefont {B.~A.}\ \bibnamefont
  {{Pazmi\~{n}o Betancourt}}}, \bibinfo {author} {\bibfnamefont {P.~Z.}\
  \bibnamefont {Hanakata}}, \bibinfo {author} {\bibfnamefont {F.~W.}\
  \bibnamefont {Starr}}, \ and\ \bibinfo {author} {\bibfnamefont {J.~F.}\
  \bibnamefont {Douglas}},\ }\href {\doibase 10.1073/pnas.1418654112}
  {\bibfield  {journal} {\bibinfo  {journal} {Proc. Natl. Acad. Sci.}\ }\textbf
  {\bibinfo {volume} {112}},\ \bibinfo {pages} {2966} (\bibinfo {year}
  {2015})}\BibitemShut {NoStop}%
\bibitem [{\citenamefont {Mirigian}\ and\ \citenamefont
  {Schweizer}(2014)}]{Mirigian2014}%
  \BibitemOpen
  \bibfield  {author} {\bibinfo {author} {\bibfnamefont {S.}~\bibnamefont
  {Mirigian}}\ and\ \bibinfo {author} {\bibfnamefont {K.~S.}\ \bibnamefont
  {Schweizer}},\ }\href {\doibase 10.1063/1.4900507} {\bibfield  {journal}
  {\bibinfo  {journal} {J. Chem. Phys.}\ }\textbf {\bibinfo {volume} {141}},\
  \bibinfo {pages} {161103} (\bibinfo {year} {2014})}\BibitemShut {NoStop}%
\bibitem [{\citenamefont {Hocky}\ \emph {et~al.}(2012)\citenamefont {Hocky},
  \citenamefont {Markland},\ and\ \citenamefont {Reichman}}]{Hocky2012}%
  \BibitemOpen
  \bibfield  {author} {\bibinfo {author} {\bibfnamefont {G.~M.}\ \bibnamefont
  {Hocky}}, \bibinfo {author} {\bibfnamefont {T.~E.}\ \bibnamefont {Markland}},
  \ and\ \bibinfo {author} {\bibfnamefont {D.~R.}\ \bibnamefont {Reichman}},\
  }\href@noop {} {\bibfield  {journal} {\bibinfo  {journal} {Phys. Rev. Lett.}\
  }\textbf {\bibinfo {volume} {108}},\ \bibinfo {pages} {225506} (\bibinfo
  {year} {2012})}\BibitemShut {NoStop}%
\bibitem [{\citenamefont {Butler}\ and\ \citenamefont
  {Harrowell}(1991)}]{Butler1991}%
  \BibitemOpen
  \bibfield  {author} {\bibinfo {author} {\bibfnamefont {S.}~\bibnamefont
  {Butler}}\ and\ \bibinfo {author} {\bibfnamefont {P.}~\bibnamefont
  {Harrowell}},\ }\href {\doibase 10.1063/1.461769} {\bibfield  {journal}
  {\bibinfo  {journal} {J. Chem. Phys.}\ }\textbf {\bibinfo {volume} {95}},\
  \bibinfo {pages} {4466} (\bibinfo {year} {1991})}\BibitemShut {NoStop}%
\bibitem [{\citenamefont {Glor}\ \emph {et~al.}(2015)\citenamefont {Glor},
  \citenamefont {Composto},\ and\ \citenamefont {Fakhraai}}]{Glor2015}%
  \BibitemOpen
  \bibfield  {author} {\bibinfo {author} {\bibfnamefont {E.~C.}\ \bibnamefont
  {Glor}}, \bibinfo {author} {\bibfnamefont {R.~J.}\ \bibnamefont {Composto}},
  \ and\ \bibinfo {author} {\bibfnamefont {Z.}~\bibnamefont {Fakhraai}},\
  }\href {\doibase 10.1021/acs.macromol.5b00979} {\bibfield  {journal}
  {\bibinfo  {journal} {Macromolecules}\ }\textbf {\bibinfo {volume} {48}},\
  \bibinfo {pages} {6682} (\bibinfo {year} {2015})}\BibitemShut {NoStop}%
\bibitem [{\citenamefont {Xie}\ \emph {et~al.}(1998)\citenamefont {Xie},
  \citenamefont {Karim}, \citenamefont {Douglas}, \citenamefont {Han},\ and\
  \citenamefont {Weiss}}]{Xie1998}%
  \BibitemOpen
  \bibfield  {author} {\bibinfo {author} {\bibfnamefont {R.}~\bibnamefont
  {Xie}}, \bibinfo {author} {\bibfnamefont {A.}~\bibnamefont {Karim}}, \bibinfo
  {author} {\bibfnamefont {J.}~\bibnamefont {Douglas}}, \bibinfo {author}
  {\bibfnamefont {C.}~\bibnamefont {Han}}, \ and\ \bibinfo {author}
  {\bibfnamefont {R.}~\bibnamefont {Weiss}},\ }\href@noop {} {\bibfield
  {journal} {\bibinfo  {journal} {Phys. Rev. Lett.}\ }\textbf {\bibinfo
  {volume} {81}},\ \bibinfo {pages} {1251} (\bibinfo {year}
  {1998})}\BibitemShut {NoStop}%
\bibitem [{\citenamefont {Farahzadi}\ \emph {et~al.}(2010)\citenamefont
  {Farahzadi}, \citenamefont {Niyamakom}, \citenamefont {Beigmohamadi},
  \citenamefont {Meyer}, \citenamefont {Keiper}, \citenamefont {Heuken},
  \citenamefont {Ghasemi}, \citenamefont {{Rahimi Tabar}}, \citenamefont
  {Michely},\ and\ \citenamefont {Wuttig}}]{Farahzadi2010}%
  \BibitemOpen
  \bibfield  {author} {\bibinfo {author} {\bibfnamefont {A.}~\bibnamefont
  {Farahzadi}}, \bibinfo {author} {\bibfnamefont {P.}~\bibnamefont
  {Niyamakom}}, \bibinfo {author} {\bibfnamefont {M.}~\bibnamefont
  {Beigmohamadi}}, \bibinfo {author} {\bibfnamefont {N.}~\bibnamefont {Meyer}},
  \bibinfo {author} {\bibfnamefont {D.}~\bibnamefont {Keiper}}, \bibinfo
  {author} {\bibfnamefont {M.}~\bibnamefont {Heuken}}, \bibinfo {author}
  {\bibfnamefont {F.}~\bibnamefont {Ghasemi}}, \bibinfo {author} {\bibfnamefont
  {M.~R.}\ \bibnamefont {{Rahimi Tabar}}}, \bibinfo {author} {\bibfnamefont
  {T.}~\bibnamefont {Michely}}, \ and\ \bibinfo {author} {\bibfnamefont
  {M.}~\bibnamefont {Wuttig}},\ }\href@noop {} {\bibfield  {journal} {\bibinfo
  {journal} {EPL (Europhysics Lett.)}\ }\textbf {\bibinfo {volume} {90}},\
  \bibinfo {pages} {10008} (\bibinfo {year} {2010})}\BibitemShut {NoStop}%
\bibitem [{\citenamefont {Zhang}\ \emph {et~al.}(2015)\citenamefont {Zhang},
  \citenamefont {Brian},\ and\ \citenamefont {Yu}}]{Zhang2015}%
  \BibitemOpen
  \bibfield  {author} {\bibinfo {author} {\bibfnamefont {W.}~\bibnamefont
  {Zhang}}, \bibinfo {author} {\bibfnamefont {C.~W.}\ \bibnamefont {Brian}}, \
  and\ \bibinfo {author} {\bibfnamefont {L.}~\bibnamefont {Yu}},\ }\href
  {\doibase 10.1021/jp5127464} {\bibfield  {journal} {\bibinfo  {journal} {J.
  Phys. Chem. B}\ }\textbf {\bibinfo {volume} {119}},\ \bibinfo {pages} {5071}
  (\bibinfo {year} {2015})}\BibitemShut {NoStop}%
\bibitem [{\citenamefont {Oron}\ and\ \citenamefont
  {Bankoff}(1997)}]{Oron1997}%
  \BibitemOpen
  \bibfield  {author} {\bibinfo {author} {\bibfnamefont {A.}~\bibnamefont
  {Oron}}\ and\ \bibinfo {author} {\bibfnamefont {S.~G.}\ \bibnamefont
  {Bankoff}},\ }\href@noop {} {\bibfield  {journal} {\bibinfo  {journal} {Rev.
  Mod. Phys.}\ }\textbf {\bibinfo {volume} {69}},\ \bibinfo {pages} {931}
  (\bibinfo {year} {1997})}\BibitemShut {NoStop}%
\bibitem [{\citenamefont {Meredith}\ \emph {et~al.}(2000)\citenamefont
  {Meredith}, \citenamefont {Smith}, \citenamefont {Karim},\ and\ \citenamefont
  {Amis}}]{Meredith2000}%
  \BibitemOpen
  \bibfield  {author} {\bibinfo {author} {\bibfnamefont {J.~C.}\ \bibnamefont
  {Meredith}}, \bibinfo {author} {\bibfnamefont {A.~P.}\ \bibnamefont {Smith}},
  \bibinfo {author} {\bibfnamefont {A.}~\bibnamefont {Karim}}, \ and\ \bibinfo
  {author} {\bibfnamefont {E.~J.}\ \bibnamefont {Amis}},\ }\href {\doibase
  10.1021/ma001298g} {\bibfield  {journal} {\bibinfo  {journal}
  {Macromolecules}\ }\textbf {\bibinfo {volume} {33}},\ \bibinfo {pages} {9747}
  (\bibinfo {year} {2000})}\BibitemShut {NoStop}%
\bibitem [{\citenamefont {Walters}\ \emph {et~al.}(2015)\citenamefont
  {Walters}, \citenamefont {Richert},\ and\ \citenamefont
  {Ediger}}]{Walters2015}%
  \BibitemOpen
  \bibfield  {author} {\bibinfo {author} {\bibfnamefont {D.~M.}\ \bibnamefont
  {Walters}}, \bibinfo {author} {\bibfnamefont {R.}~\bibnamefont {Richert}}, \
  and\ \bibinfo {author} {\bibfnamefont {M.~D.}\ \bibnamefont {Ediger}},\
  }\href {\doibase 10.1063/1.4916649} {\bibfield  {journal} {\bibinfo
  {journal} {J. Chem. Phys.}\ }\textbf {\bibinfo {volume} {142}},\ \bibinfo
  {pages} {134504} (\bibinfo {year} {2015})}\BibitemShut {NoStop}%
\bibitem [{\citenamefont {Liu}\ \emph {et~al.}(2015{\natexlab{b}})\citenamefont
  {Liu}, \citenamefont {Cheng}, \citenamefont {Salami}, \citenamefont {Gao},
  \citenamefont {Glor}, \citenamefont {Li}, \citenamefont {Walsh},\ and\
  \citenamefont {Fakhraai}}]{Liu2015}%
  \BibitemOpen
  \bibfield  {author} {\bibinfo {author} {\bibfnamefont {T.}~\bibnamefont
  {Liu}}, \bibinfo {author} {\bibfnamefont {K.}~\bibnamefont {Cheng}}, \bibinfo
  {author} {\bibfnamefont {E.}~\bibnamefont {Salami}}, \bibinfo {author}
  {\bibfnamefont {F.}~\bibnamefont {Gao}}, \bibinfo {author} {\bibfnamefont
  {E.}~\bibnamefont {Glor}}, \bibinfo {author} {\bibfnamefont {M.}~\bibnamefont
  {Li}}, \bibinfo {author} {\bibfnamefont {P.}~\bibnamefont {Walsh}}, \ and\
  \bibinfo {author} {\bibfnamefont {Z.}~\bibnamefont {Fakhraai}},\ }\href
  {\doibase 10.1039/C5SM01044F} {\bibfield  {journal} {\bibinfo  {journal}
  {Soft Matter}\ }\textbf {\bibinfo {volume} {11}},\ \bibinfo {pages} {7558}
  (\bibinfo {year} {2015}{\natexlab{b}})}\BibitemShut {NoStop}%
\bibitem [{\citenamefont {Hanakata}\ \emph {et~al.}(2014)\citenamefont
  {Hanakata}, \citenamefont {Douglas},\ and\ \citenamefont
  {Starr}}]{Hanakata2014}%
  \BibitemOpen
  \bibfield  {author} {\bibinfo {author} {\bibfnamefont {P.~Z.}\ \bibnamefont
  {Hanakata}}, \bibinfo {author} {\bibfnamefont {J.~F.}\ \bibnamefont
  {Douglas}}, \ and\ \bibinfo {author} {\bibfnamefont {F.~W.}\ \bibnamefont
  {Starr}},\ }\href {\doibase 10.1038/ncomms5163} {\bibfield  {journal}
  {\bibinfo  {journal} {Nat. Commun.}\ }\textbf {\bibinfo {volume} {5}},\
  \bibinfo {pages} {4163} (\bibinfo {year} {2014})}\BibitemShut {NoStop}%
\bibitem [{\citenamefont {Stevenson}\ and\ \citenamefont
  {Wolynes}(2008)}]{Stevenson2008}%
  \BibitemOpen
  \bibfield  {author} {\bibinfo {author} {\bibfnamefont {J.~D.}\ \bibnamefont
  {Stevenson}}\ and\ \bibinfo {author} {\bibfnamefont {P.~G.}\ \bibnamefont
  {Wolynes}},\ }\href {\doibase 10.1063/1.3041651} {\bibfield  {journal}
  {\bibinfo  {journal} {J. Chem. Phys.}\ }\textbf {\bibinfo {volume} {129}},\
  \bibinfo {pages} {234514} (\bibinfo {year} {2008})}\BibitemShut {NoStop}%
\end{thebibliography}%

\end{document}